\documentclass[aps,pra,reprint, amsmath, amssymb,superscriptaddress]{revtex4-1}

\usepackage{bm}
\usepackage[retainorgcmds]{IEEEtrantools}
\usepackage{graphicx}
\usepackage{mathrsfs}
\usepackage{amsmath}
\usepackage{amssymb}
\usepackage{color}
\usepackage{amsfonts}
\usepackage{times,txfonts}
\usepackage{nicefrac}

\newcommand{\ud}{\,\mathrm{d}}

\begin{document}

\title{Lindblad-Floquet description of finite-time quantum heat engines}
\date{\today}
\author{Stefano Scopa}
\affiliation{
Laboratoire de Physique et Chimie Th\'eoriques, UMR CNRS 7019, Universit\'e de Lorraine BP 239 F-54506 Vandoeuvre-l\'es-Nancy Cedex, France
}
\author{Gabriel T. Landi}
\email{gtlandi@if.usp.br}
\affiliation{Instituto de F\'isica da Universidade de S\~ao Paulo,  05314-970 S\~ao Paulo, Brazil}
\author{Dragi Karevski}
\affiliation{
Laboratoire de Physique et Chimie Th\'eoriques, UMR CNRS 7019, Universit\'e de Lorraine BP 239 F-54506 Vandoeuvre-l\'es-Nancy Cedex, France
}

\begin{abstract}

The operation of autonomous finite-time quantum heat engines rely on the existence of a stable limit cycle in which the dynamics becomes periodic. 
The two main questions that naturally arise are therefore whether such a limit cycle will eventually be reached and, once it has, what is the state of the system within the limit cycle. 
In this paper we show that the application of Floquet's theory to Lindblad dynamics offers clear answers to both questions. 
By moving to a generalized rotating frame, we show that it is possible to identify a single object,  the Floquet Liouvillian, which encompasses all operating properties of the engine. 
First, its spectrum dictates the convergence to a limit cycle.
And second, the state within the limit cycle is  precisely its zero eigenstate, therefore reducing the problem to that of determining the steady-state of a time-independent master equation. 
To illustrate the usefulness of this theory, we apply it to a harmonic oscillator subject to a time-periodic  work protocol and time-periodic   dissipation, an open-system generalization of the  Ermakov-Lewis theory.
The use of this theory to implement a finite-time Carnot engine subject to continuous frequency modulations is also discussed. 

\end{abstract}
\maketitle{}

%
%
%
%

\section{\label{sec:int}Introduction}

The last decades have witnessed  remarkable progress in the experimental manipulation of a variety of quantum platforms, enabling for the first time the coherent control over genuinely quantum mechanical resources. 
This progress has in turn motivated substantial research in addressing which types of applications may be drawn up using such platforms. 
One  promising such avenue concerns the use of quantum effects in the operation of heat engines and refrigerators \cite{Alicki1979,Rezek2006,Kosloff2014,Abah2012,Alicki2015}. 
This would allow, for instance, to extend the efficiency above  classical bounds using non-equilibrium reservoirs \cite{Abah2014,Klaers2017a,Ronagel2014,Correa2014,Jaramillo2016,Samuelsson2017}, operate adiabatic cycles  at finite times using shortcuts to adiabaticity \cite{DelCampo2014,Abah2016,Abah2017}, implement informationally driven engines  \cite{Camati2016,Elouard2017,Cottet2017,Masuyama2017,Manzano2017b} and even  substitute work and heat by quantum resources, such as entanglement \cite{Micadei2017} and coherence \cite{Manzano2017}. 

Despite the surge in interest, several scenarios still remain  unexplored, with most studies so far having focused on either continually operated engines, such as absorption refrigerators \cite{Kosloff2014,Kilgour2018,Latune2018,Schilling,Holubec2018,Hofer2018} or stroke-based engines, with a particular focus on the Otto cycle
\cite{Alicki1979,Rezek2006,Kosloff2014,Abah2012,Alicki2015,Abah2014,Klaers2017a,Ronagel2014,Correa2014,Jaramillo2016,Samuelsson2017,DelCampo2014,Abah2016,Abah2017,Insinga2018,Insinga2016,Feldmann1996,Feldmann2004,Rezek2006a,Watanabe2017,Kieu2004,Song2016,Altintas2015}.
Carnot or Stirling cycles have also been studied, but to a lesser extent \cite{He2002,Quan2009,Gardas2015,Brandner2016}.  
The same is true for continually operated work protocols \cite{Alicki2006,Alicki2012,Kosloff2013,Szczygielski2013}.

The main reason behind the focus on the Otto cycle is due to  its convenient separation of the work and heat strokes, which facilitates the theoretical modeling since it allows one to use  unitary dynamics for the former  and time-independent dissipative dynamics for the later. 
Moreover, it is usually assumed that the heat strokes  act for a sufficiently long time so as to allow for a full thermalization.
Studies dealing with finite-time operations over all strokes remain scarce.
In this case a subtle question arises concerning the convergence or not to a limit-cycle, in which the operation of the engine becomes periodic.
As shown in \cite{Insinga2018,Insinga2016,Feldmann1996,Feldmann2004,Watanabe2017}, depending on the work protocol and its corresponding injection of energy, unusually large values of dissipation may be necessary in order to obtain a stable cycle. 
Consequently, these considerations may have an important impact in translating optimization protocols, such as shortcuts to adiabaticity, to the finite-time regime. 

In the context of finite-time engines, the problem may therefore be divided in two parts. 
The first is the convergence towards a limit cycle and the second concerns the behavior of the system within this cycle. 
A more thorough understanding of these two features is therefore essential for advancing our theoretical understanding of quantum heat engines. 
However, such advances are hampered by the lack of a consistent theoretical framework to address the properties of the cycle as a whole, instead of  each individual stroke.

In this paper we attempt to fulfill this gap by showing how the operation of a heat engine may be neatly formulated in terms of Floquet's theory applied to Lindblad dynamics. 
Instead of analyzing each stroke separately, we consider the full cycle as modeled by a time-dependent  periodic Liouvillian $\mathcal{L}_t$. 
Then, applying Floquet's theory and moving to a generalized rotating frame, we show how all relevant aspects of the heat engine are completely dictated by a super-operator $\mathcal{L}_F(t)$, which we  refer to as Floquet Liouvillian. 
First, the convergence to a limit cycle is directly associated to the eigenvalues of $\mathcal{L}_F(t)$ (which are independent of $t$). 
Second, the state of the system after the limit-cycle has been reached is precisely the zero-eigenstate of $\mathcal{L}_F(t)$, with $t$ as a parameter. 
This therefore reduces the problem of finding the limit cycle to that of finding the steady-state of a \emph{time-independent} master equation. 

Floquet theory has seen a boom of interest in the last decades, specially due to its potential use in quantum simulation. 
Important advances in its extension to open quantum systems have also appeared recently \cite{Haddadfarshi2015,Wu2015a,Restrepo2016,Hartmann2016b,Dai2016,Alicki2006,Alicki2012}. 
In particular, we call attention to Refs.~\cite{Alicki2006,Alicki2012} where the authors describe a method to derive quantum master equations for systems subject to a periodically driven Hamiltonian and in contact with a heat bath. 
Such a framework has direct  applications in the context of quantum heat engines (c.f.~\cite{,Szczygielski2013,Kosloff2014}).
However, it cannot be used to describe stroke-based engines containing unitary (isentropic) branches, such as the Carnot cycle, since it assumes that the system-bath coupling  remains on at all times. 
To do so,  one must be able to couple and uncouple the system from the bath periodically. 
Although there are ways of bypassing this, such as using sufficiently fast drives or implementing protocols which ensure that the heat flow is zero on average   \cite{Martinez2015a,Martinez2015}, in order to describe Carnot and related cycles in all generality, one must eventually make use of some uncoupling mechanism. 
This is the main advantage behind our approach. 
We shall take as a starting point an arbitrary master equation, but whose parameters are assumed to vary periodically following some protocol (while ensuring complete positivity at all times). 
Even though this has the disadvantage of loosing the microscopic interpretation of a system-environment coupling, it has the advantage of being able to deal with arbitrary coupling protocols. 
Moreover, it also works for engineered reservoirs and phenomenological equations.

To illustrate the usefulness of this theory, we  apply it to the exactly soluble model of a harmonic oscillator under the influence of an arbitrary time-periodic  work protocol and arbitrary time-periodic and Gaussian-preserving  environments. 
This corresponds to an open-system  generalization of the   Ermakov-Lewis theory \cite{Lewis1967,Lewis1968,Jeremie2010,Scopa2018} describing a harmonic oscillator subject to a  frequency modulation.
In our theory all parameters may be time-dependent, including the frequency, mass, damping rate, temperature and squeezing (magnitude and angle), provided they all share the common period of the cycle. 
This therefore allows one to implement any type of single-oscillator engine,  including protocols for shortcuts to adiabaticity or the use of (possibly time-dependent) squeezing effects to maximize efficiency.  
Given the widespread use of harmonic systems as working fluids, we believe that these results should prove valuable in the design and optimization of more efficient engines. 
As an example, we briefly discuss their application to the study of a finite-time Carnot engine.

%
%
%
%

\section{\label{sec:theory}Lindblad-Floquet  theory}

We consider here an arbitrary quantum heat engine operating periodically with period $\mathcal{T}$. 
The different strokes of the engine may contain both unitary and dissipative contributions. 
Instead of describing these effects separately, we combine them into a single master equation for the working fluid's density matrix $\rho(t)$, subject to a certain time-dependent periodic Liouvillian operator $\mathcal{L}_t$ satisfying $\mathcal{L}_{t+\mathcal{T}} = \mathcal{L}_t$.
For convenience we write the master equation in super-operator space as 
\begin{equation}\label{M}
\frac{\ud |\rho_t\rangle}{\ud t} = \mathcal{L}_t |\rho_t\rangle,
\end{equation}
where $|\rho_t\rangle$ is the density matrix written in vectorized form.

\begin{figure}[!t]
\centering
\qquad
\includegraphics[width=0.5\textwidth]{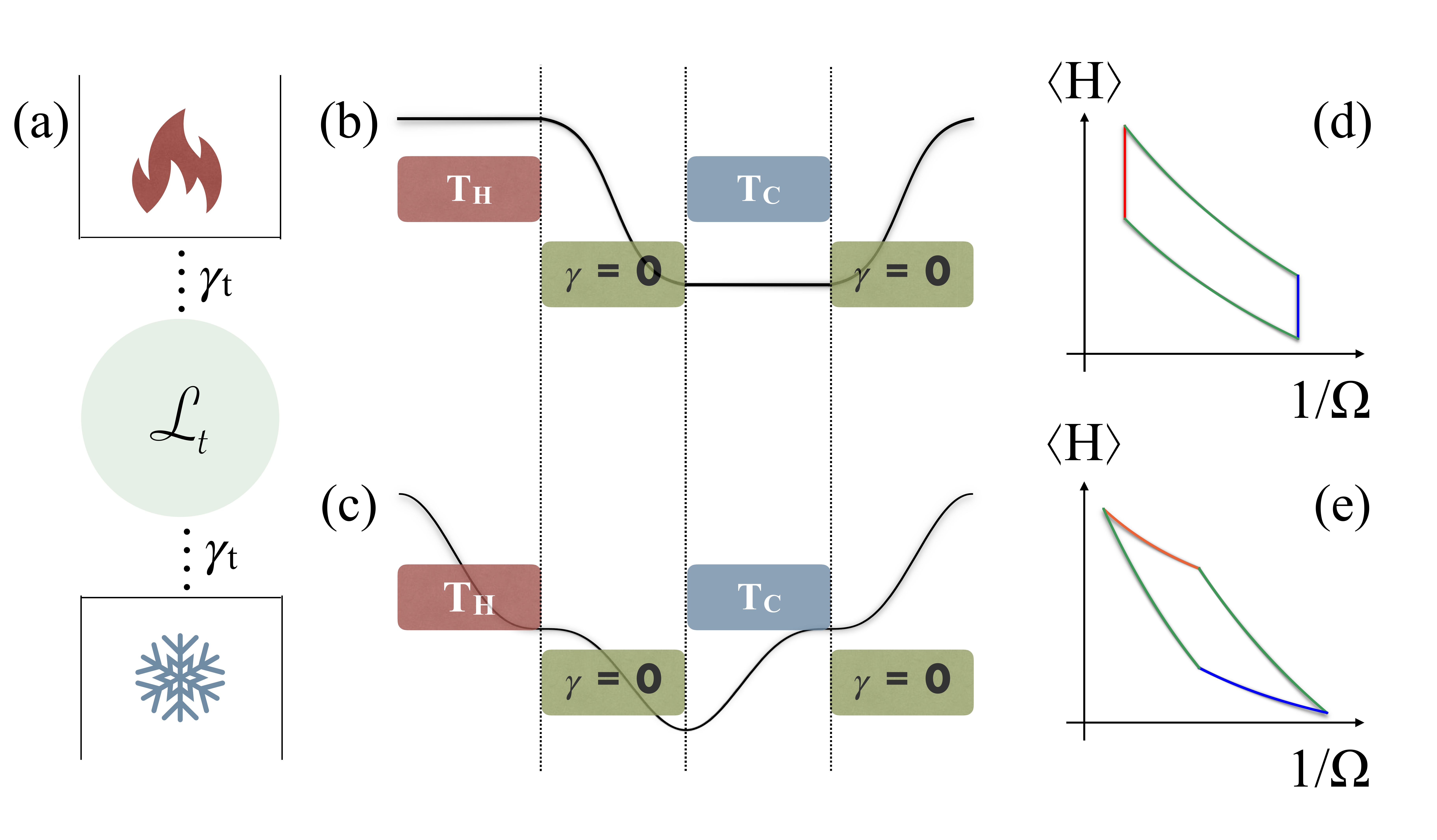}
\caption{\label{fig:drawing}
Modeling of quantum heat engines using Lindblad-Floquet theory. 
(a) A thermodynamic cycle, for instance with a harmonic oscillator as working fluid, may be described by means of a time-periodic Liouvillian $\mathcal{L}_t$ which encompasses the evolution of the work parameter $\Omega(t)$, the different bath temperatures $T_H$ and $T_C$, and the on-off protocol for the bath couplings, modeled by $\gamma_t$.
(b,c) Diagrams for the Otto and Carnot cycles respectively. 
The green regions correspond to the unitary expansion and compression strokes, whereas the red and blue regions denote the interactions with the two reservoirs. 
The black curves illustrate typical work protocols $\Omega(t)$ for the two cycles. 
In the Otto cycle $\Omega$ is constant in the red and blue regions, whereas in the Carnot cycle the hot isothermal is an expansion and the cold isothermal a compression. 
(d,e) The corresponding quasi-static cycles in a $\langle H \rangle$~vs.~$1/\Omega$ diagram.
}
\end{figure}

To clarify our notation, let us consider the implementation of an Otto cycle. 
Let $\Omega(t)$ denote a generic work parameter, $\gamma(t)$ the bath-coupling constant and $T(t)$ the temperature. 
An Otto cycle would then be described by the following protocol (see  Fig.~\ref{fig:drawing}(b)):
\begin{IEEEeqnarray*}{rCl}
\Omega: & \quad & (\Omega_1, \Omega_2(t), \Omega_3, \Omega_4(t)), \\[0.2cm]
\gamma: & \quad & (\gamma_1,0,\gamma_3,0), \\[0.2cm]
T: &\quad & (T_H, -, T_C, -).
\end{IEEEeqnarray*}
The first and third strokes are the thermalization (isochoric) heat strokes for which the work parameter is fixed and the system is allowed to partially  relax in contact with heat baths at temperatures $T_H$ and $T_C$. 
The second and fourth strokes, on the other hand, are the unitary  (isentropic) work strokes for which the system is detached from the bath ($\gamma = 0$). 
Each stroke may have different durations but we denote by $\mathcal{T}$ the total period of the cycle.
It is clear from this example that any type of cycle  may be constructed by appropriately choosing the parameters in the master equation. 
For instance, the protocol for constructing a Carnot cycle is illustrated in Fig.~\ref{fig:drawing}(c).

We now cast the master Eq.~(\ref{M}) within the framework of Floquet's theory. 
We begin by moving to a generalized rotating frame by defining a new state $|\tilde{\rho}_t\rangle = W(t) |\rho_t\rangle$, where $W(t)$ is a time-periodic super-operator.  
In this rotating frame $|\tilde{\rho}_t\rangle$ will satisfy an equation analogous to~(\ref{M}), but subject to the effective Liouvillian 
\begin{equation}\label{Lt}
\tilde{\mathcal{L}} = W(t) \mathcal{L}_t W^{-1} (t) + \frac{\ud W(t)}{\ud t} W^{-1}(t).
\end{equation}
If we can  now choose $W(t)$ such that $\tilde{\mathcal{L}}$ is time-independent, then the evolution in this rotating  frame, between times $t_0$ and $t$, will be given simply by $|\tilde{\rho}_t \rangle = e^{(t-t_0)\tilde{\mathcal{L}}} |\tilde{\rho}_{t_0}\rangle$. 
Moving back to the original frame then allows us to write the evolution of $|\rho_t\rangle$ as 
\begin{equation}
|\rho_t\rangle = K(t,t_0) e^{(t-t_0) \mathcal{L}_F(t_0)} |\rho_{t_0}\rangle,
\end{equation}
where we have defined the Floquet Liouvillian  
\begin{equation}\label{LF}
\mathcal{L}_F(t_0) = W^{-1}(t_0) \tilde{\mathcal{L}} W(t_0),
\end{equation}
and the micromotion super-operator 
\begin{equation}\label{K}
K(t,t_0) = W^{-1}(t) W(t_0).
\end{equation}
Since $W(t)$ is periodic, it follows that $\mathcal{L}_F(t_0+\mathcal{T}) = \mathcal{L}_F(t_0)$. 
The same is   true for both arguments of $K(t,t_0)$.

From Eq.~(\ref{K}) one also has that $K(t_0,t_0) = 1$. 
Hence, the stroboscopic evolution of the system will be governed solely by the Floquet Liouvillian 
\begin{equation}\label{strobo}
|\rho_{t_0 + n\mathcal{T}}\rangle = e^{n \mathcal{T} \mathcal{L}_F(t_0)} |\rho_{t_0}\rangle.
\end{equation}
This leads us to our first main result: 
\emph{A sufficient condition for the system to converge to a limit cycle is that all eigenvalues of $\mathcal{L}_F(t_0)$  have a non-positive real part.} 
Instead of looking at $\mathcal{L}_F(t_0)$,  one may also look directly at $\tilde{\mathcal{L}}$ since the two are connected by a similarity transformation [Eq.~(\ref{LF})].
As will be illustrated below, this will in general be much simpler and also serves to show that the spectrum of $\mathcal{L}_F(t_0)$ is independent of $t_0$.


Next let $|\rho_F(t_0)\rangle$ denote the steady-state of $\mathcal{L}_F(t_0)$; that is, $\mathcal{L}_F(t_0) |\rho_F(t_0) \rangle = 0$. 
For simplicity, we will  henceforth assume that this steady-state is unique. 
Then, after a sufficiently long time, the engine will  eventually converge to a limit cycle for which  the density matrix becomes periodic, being given by:
\begin{equation}\label{cycle}
|\rho_t\rangle = K(t,t_0) |\rho_F(t_0)\rangle.
\end{equation}
However, since the micromotion is periodic, we may now set $t_0 = t$, which then finally gives 
\begin{equation}
|\rho_t\rangle  = |\rho_F(t)\rangle.
\end{equation}
This is our second main result:  \emph{within the limit-cycle the state of the system will be simply the zero eigenstate of $\mathcal{L}_F(t)$, provided this eigenstate is unique}.
This result is a consequence of the convergence towards a steady-state.
It is therefore a feature  unique of open systems and allows for a remarkable simplification in the description of the problem. 

It is also useful to compare these results with the method developed in Refs.~\cite{Alicki2006,Alicki2012}, which consists in generalizing the microscopic derivations to include a periodic drive in the system Hamiltonian. 
Our approach can therefore be viewed as complementary.
We assume no information  about the environment or the processes which led to the master equation. 
Instead, we take the Liouvillian as given and then cast it in terms of Floquet's theory to extract its main properties.
This has the advantage of allowing for the coupling constant and the bath temperature to be turned on and off at will. 
One should note, however, that even if the bath parameters are constant, both models may not necessarily give the same result. 
The reason is that in the method of Refs.~\cite{Alicki2006,Alicki2012} one takes into account the effects that the driving on the system have on the exchange of excitations between system and bath. 
However, it is expected that these effects will become important only if the time-scales of the drive become comparable with the bath correlation times. 
If that is not the case, then from physical grounds one expects that the master equation will be such that it instantaneously thermalize the system at each instant of time.

We also would like to call attention to some known difficulties of dealing with Floquet Liouvillians. 
Even though the computation of the stroboscopic map~(\ref{strobo}) is always well defined, the calculation of the Floquet Liouvillian $\mathcal{L}_F$ may be problematic, leading to generators that do not preserve complete positivity. 
A numerically exact illustration of this was given recently in Fig.~4 of Ref.~\cite{Hartmann2016b}.
This can turn out to be a serious issue when one is interested in finding $\mathcal{L}_F$ by means of high-frequency Magnus expansions, 
which is often the case since  the problem is usually analytically intractable. 
In this sense, another important development to call attention to is  Ref.~\cite{Haddadfarshi2015},  where the authors have developed a method to build high-frequency Magnus expansions that preserve complete positivity at all orders.
Interestingly, the seeming inconsistency between this and the results in Ref.~\cite{Hartmann2016b}, seem to point to a limited radius of convergence of the Magnus expansion. 
 Here we shall avoid this issue by looking at an  exactly soluble model for which the dynamics is always completely positive.


%
%
%
%

\section{\label{sec:app}Application to a harmonic oscillator}

We now consider the exactly soluble model of a bosonic mode, described by an annihilation operator $a$, subject to an arbitrary time-dependent and  Gaussian-preserving open system dynamics. 
Here we provide only the main ideas and results,  leaving some of the technical details to the appendices. 
The Hamiltonian of the system is chosen to be
\begin{equation}\label{H_QHO}
H_t = \omega_t (a^\dagger a+\nicefrac{1}{2}) + \frac{\lambda_t}{2} a a + \frac{\lambda_t^*}{2} a^\dagger a^\dagger.
\end{equation}
where $\omega_t$ and $\lambda_t$ are arbitrary periodic functions satisfying  $\omega_t^2 > |\lambda_t|^2$.
In a mechanical picture, the Hamiltonian~(\ref{H_QHO}) describes a situation where both the mass and the spring constant may be time-dependent. 
The situation where the mass is constant corresponds to $\eta:= \omega_t - \lambda_t$ being time-independent, in which case the mechanical frequency $\Omega_t$ is given by  $\Omega_t^2 = \omega_t^2 - \lambda_t^2$ (see Appendix~\ref{sec:relation}).

The Hamiltonian~(\ref{H_QHO}) adds to $\mathcal{L}_t$  three super-operators: 
\begin{IEEEeqnarray}{rCl}
\label{H0}
\mathcal{H}_0 &=& - i [a^\dagger a, \bullet],	\\[0.2cm]
\label{H1}
\mathcal{H}_1 &=& - i [a a, \bullet], 	\\[0.2cm]
\label{H2}
\mathcal{H}_2 &=& - i [a^\dagger a^\dagger, \bullet].
\end{IEEEeqnarray}
In addition, we consider the general effects of Gaussian preserving dissipation generated by
\begin{IEEEeqnarray}{rCl}
\mathcal{D}_1 = a \bullet a^\dagger - \frac{1}{2} \{a^\dagger a, \bullet\}, 
&\qquad&
\mathcal{D}_2 = a^\dagger \bullet a - \frac{1}{2} \{ a a^\dagger, \bullet\},
\IEEEeqnarraynumspace	\\[0.2cm]
\mathcal{D}_3 = a^\dagger \bullet a^\dagger - \frac{1}{2} \{ a^\dagger a^\dagger, \bullet\},	
&\qquad&
\mathcal{D}_4 =  a \bullet a - \frac{1}{2} \{ a a, \bullet\}.\IEEEeqnarraynumspace
\end{IEEEeqnarray}
With these ingredients, we then parametrize our  time-dependent Liouvillian as 
\begin{equation}\label{L}
\mathcal{L}_t = \mathcal{H}_t + \mathcal{D}_t,
\end{equation}
where 
\begin{IEEEeqnarray}{rCl}
\label{Ht}
\mathcal{H}_t &=& \omega_t \mathcal{H}_0 + \frac{\lambda_t}{2} \mathcal{H}_1 + \frac{\lambda_t^*}{2} \mathcal{H}_2,	\\[0.2cm]
\label{Dt}
\mathcal{D}_t &=& \gamma_t (N_t +1) \mathcal{D}_1 + \gamma_t N_t \mathcal{D}_2  - \gamma_t M_t \mathcal{D}_3 - \gamma_t M_t^* \mathcal{D}_4.
\end{IEEEeqnarray}
where $\gamma_t$, $N_t$ and $M_t$ are periodic parameters satisfying $\gamma_t>0$ and $N_t (N_t + 1) > |M_t|^2$.
Here $\gamma_t$ represents the coupling strength to the bath, whereas $N_t$ and $M_t$ may represent both thermal  and squeezing effects, depending on the choice of basis. 
For instance, if $\lambda_t =0$ then  a thermal bath at a temperature $T$ correspond to $M_t =0$ and $N_t = (e^{\omega_t/T}-1)^{-1}$.
For the general Hamiltonian~(\ref{H_QHO}), on the other hand, the thermal bath is modeled  by 
\begin{equation}\label{isothermal}
N_t+ \nicefrac{1}{2} = \frac{\omega_t}{2\Omega_t} \coth\bigg(\frac{\Omega_t}{2T}\bigg),
\quad
M_t =  -\frac{\lambda_t}{2\Omega_t} \coth\bigg(\frac{\Omega_t}{2T}\bigg),
\end{equation}
where $\Omega_t^2 = \omega_t^2 - \lambda_t^2$.

Next we apply the rotating frame transformation~(\ref{Lt}).
The key property making this problem analytically tractable and free of the aforementioned positivity issues is that the  7 super-operators $\{\mathcal{H}_i,\mathcal{D}_i\}$ form a closed algebra \cite{PeixotoDeFaria2007} (see also  \cite{Ryabov2013}).
In particular, the sets $\{\mathcal{H}_i\}$ and $\{\mathcal{D}_i\}$,  when taken separately,  satisfy independent algebras: 
\begin{IEEEeqnarray}{rCl}
[\mathcal{H}_0, \mathcal{H}_{1,2}] &=& \pm2 i \mathcal{H}_{1,2}, 	
\qquad
[\mathcal{H}_1, \mathcal{H}_2] = -4 i \mathcal{H}_0, 	
\end{IEEEeqnarray}
and
\begin{IEEEeqnarray}{rCl}
[\mathcal{D}_1, \mathcal{D}_2] &=& -(\mathcal{D}_1 + \mathcal{D}_2),	
\qquad
[\mathcal{D}_3, \mathcal{D}_4] = 0,	
\nonumber \\ 
&&\\
\nonumber 
[\mathcal{D}_1, \mathcal{D}_{3,4}] &=& -\mathcal{D}_{3,4},	
\qquad
[\mathcal{D}_2, \mathcal{D}_{3,4}] = \mathcal{D}_{3,4},
\end{IEEEeqnarray}
Mixtures of the two sets, on the other hand, only produce elements of the latter:
\begin{IEEEeqnarray}{rCLccrl}
[\mathcal{H}_0, \mathcal{D}_{1,2}] &=& 0,
&\qquad&	
[\mathcal{H}_0, \mathcal{D}_{3,4}] &=& \mp 2 i \mathcal{D}_{3,4}, 	
\IEEEeqnarraynumspace\\[0.2cm]
[\mathcal{H}_1, \mathcal{D}_{1,2}] &=& -2 i \mathcal{D}_4, 	
&\qquad&
[\mathcal{H}_2, \mathcal{D}_{1,2}] &=& 2 i \mathcal{D}_3, 	\\[0.2cm]
[\mathcal{H}_1, \mathcal{D}_3] &=& -2 i (\mathcal{D}_1+\mathcal{D}_2), 	
&\qquad&
[\mathcal{H}_1, \mathcal{D}_4] &=& 0, 	\\[0.2cm]
[\mathcal{H}_2, \mathcal{D}_4] &=& 2 i (\mathcal{D}_1+\mathcal{D}_2) ,
&\qquad&
[\mathcal{H}_2, \mathcal{D}_3] &=& 0.
\end{IEEEeqnarray}
This algebraic structure suggests that the operator $W(t)$ in Eq.~(\ref{Lt}) may be taken as  
\begin{equation}\label{W_app}
W(t) = V(t) U(t),
\end{equation}
where  
\begin{equation}\label{V_app}
V(t) = e^{g_1 \mathcal{D}_1} e^{g_2 \mathcal{D}_2} e^{g_3 \mathcal{D}_3} e^{g_4 \mathcal{D}_4},
\end{equation}
and 
\begin{equation}\label{U_app}
U(t) =  e^{r_0 \mathcal{H}_0}  e^{r_1 \mathcal{H}_1}  e^{r_2 \mathcal{H}_2}.
\end{equation}
Here $r_i(t)$ and $g_i(t)$ are time-periodic c-number functions  which are to be suitably adjusted so as to make $\tilde{\mathcal{L}}$ time-independent.
The problem is then solved sequentially. 
First one applies $U(t)$ and adjusts the  $r_i(t)$  to make the unitary part time-independent.
Then  $V(t)$ is applied and the $g_i(t)$ are adjusted to deal with the dissipative part.
In this section, we shall illustrate the procedure in the simpler case when $\omega = \lambda = 0$. 
That is, when only the dissipative terms are present.
The general formulation is presented in Appendices~\ref{sec:gen_uni} and \ref{sec:gen_diss} and the main results will be summarized in Sec.~\ref{ssec:app_gen} below. 

\subsection{\label{ssec:purely}Purely dissipative case}

In the case $\omega = \lambda = 0$ the situation simplifies dramatically since only the dissipative part remains in the Liouvillian~(\ref{L}). 
Consequently, it suffices to choose $U(t) = 1$ in Eq.~(\ref{W_app}). 
To carry out the rotating frame transformation in Eq.~(\ref{Lt}) it is necessary to evaluate products such as 
\begin{equation}
e^{g_1 \mathcal{D}_1} \mathcal{D}_2 e^{-g_1 \mathcal{D}_1} = e^{-g_1} \mathcal{D}_2 + (e^{-g_1} - 1) \mathcal{D}_1,
\end{equation}
which can be found as usual, with the Baker-Campbell-Haussdorff formula. 
One also requires the identity
\begin{equation}
\frac{\ud (e^{g_i \mathcal{D}_i})}{\ud t} (e^{-g_i \mathcal{D}_i}) = \frac{\ud g_i}{\ud t} \mathcal{D}_i.
\end{equation}
Carrying out all computations we then find 
\begin{equation}\label{Lt6}
\tilde{\mathcal{L}} =  C_1(t) \mathcal{D}_1 + C_2(t) \mathcal{D}_2 + C_3(t) \mathcal{D}_3 + C_4(t) \mathcal{D}_4,
\end{equation}
where 
\begin{IEEEeqnarray}{rCl}
\label{diss_C2}
C_2(t) &=& e^{-g_1} \bigg[ \dot{g}_2 - \gamma_t + \gamma_t e^{g_2} (N_t+1)\bigg],	\\[0.2cm]
\label{diss_C1}
C_1(t) &=& \dot{g}_1 - \dot{g}_2 + \gamma_t + C_2(t),		\\[0.2cm]
\label{diss_C3}
C_3(t) &=& e^{g_2 - g_1} \bigg[ \dot{g}_3 + \gamma_t g_3 - \gamma_t M_t \bigg], 	\\[0.2cm]
\label{diss_C4} 
C_4(t) &=& e^{g_2 - g_1} \bigg[ \dot{g}_4 + \gamma_t  g_4 - \gamma_t M_t^* \bigg].
\end{IEEEeqnarray}

We now must choose \emph{time-periodic} functions for the $g_i(t)$ which will make all $C_i(t)$ time-independent. 
We see that this may be accomplished by setting $g_2$, $g_3$ and $g_4$ to be the time-periodic solutions of 
\begin{IEEEeqnarray}{rCl}
\label{diss_eq2}
\dot{g}_2 - \gamma_t + \gamma_t e^{g_2} (N_t+1) &=& 0 ,		\\[0.2cm]
\label{diss_eq3}
\dot{g}_3 + \gamma_t  g_3 - \gamma_t M_t &=& 0, \\[0.2cm]
\label{diss_eq4}
\dot{g}_4 + \gamma_t  g_4 - \gamma_t M_t^* &=& 0 ,
\end{IEEEeqnarray}
which then imply $C_2 = C_3 = C_4 = 0$.
Note also that $g_4 = g_3^*$. 
Finally, in order to make $C_1(t)$ time-independent, we may choose
\begin{equation}\label{diss_eq1}
g_1(t) = g_2(t) + \int\limits_0^t \ud t' \bigg[\bar{\gamma} - \gamma(t')\bigg]	
\end{equation}
where we have defined the time-average
\begin{equation}\label{time_average}
\overline{\gamma} = \frac{1}{T} \int\limits_0^T \ud t \; \gamma(t)
\end{equation}
With this form for $g_1$ we then get $C_1 = \bar{\gamma}$ so that the rotating frame Liouvillian becomes simply 
\begin{equation}\label{diss_Lt_final}
\tilde{\mathcal{L}} = \bar{\gamma} \mathcal{D}_1
\end{equation}
Thus, in the rotating frame the system evolves as if coupled to a zero-temperature bath with damping rate $\bar{\gamma}$.

We see from Eq.~(\ref{diss_eq3}) that $g_3$ satisfies a linear differential equation, whereas the same is not true for $g_2$. 
However, if we change variables to 
\begin{equation}\label{G2_def}
G_2 = e^{- g_2} - \nicefrac{1}{2}
\end{equation}
then Eq.~(\ref{diss_eq2}) becomes
\begin{equation}\label{diss_eqG2}
\dot{G_2} + \gamma_t G_2 = \gamma_t (N_t +  \nicefrac{1}{2})
\end{equation}
which is linear in $G_2$. 
Thus, we conclude that \emph{in the case of purely dissipative dynamics, all Floquet variables satisfy linear differential equations}. 
We will see that when $\lambda_t \neq 0$ in Eq.~(\ref{H_QHO}), this will no longer be the case. 

Having found the functions which make $\tilde{\mathcal{L}}$ time-independent, we now apply the inverse procedure and compute the Floquet Liouvillian in Eq.~(\ref{LF}), 
with $\tilde{\mathcal{L}}$ being given by Eq.~(\ref{diss_Lt_final}).
As a result, we find 
\begin{equation}\label{diss_LF}
\mathcal{L}_F(t) = \bar{\gamma}(N_F(t) + 1) \mathcal{D}_1 + \bar{\gamma} N_F(t) \mathcal{D}_2 - \bar{\gamma} M_F(t) \mathcal{D}_3 - \bar{\gamma} M_F^*(t) \mathcal{D}_4.
\end{equation}
which has the same form as the original dissipator~(\ref{Dt}), but with time-independent damping $\bar{\gamma}$ and new parameters $N_F(t)$ and $M_F(t)$, which turn out to be simply given by $N_F(t) = G_2(t)$ and $M_F(t) = g_3(t)$.
Thus, in view of Eqs.~(\ref{diss_eqG2}) and (\ref{diss_eq3}), we may recast the final result as the statement that the Floquet parameters are the time-periodic solutions of 
\begin{IEEEeqnarray}{rCl}
\label{diss_NF}
\dot{N}_F + \gamma_t N_F &=& \gamma_t N_t,	\\[0.2cm]
\dot{M}_F + \gamma_t M_F &=& \gamma_t M_t.
\label{diss_MF}
\end{IEEEeqnarray}
These solutions then determine the value of the thermal noise and squeezing at any time $t$ during the limit cycle. 

In the Floquet Liouvillian the time-dependence enters only as a parameter and all we require is the steady-state of $\mathcal{L}_F$ for a given $t$. 
This state turns out to be simply a squeezed thermal state with covariances
\begin{equation}
\langle a^\dagger a \rangle_t = N_F(t),
\qquad
\langle a a \rangle_t = M_F(t).
\end{equation}
Thus, once the periodic solutions of Eqs.~(\ref{diss_NF}) and (\ref{diss_MF}) are found, one knows exactly the density matrix in the limit-cycle.

\subsection{\label{ssec:app_gen}Summary of results for the general case}

When $\lambda_t \neq 0$ the situation becomes much more complicated.
In this case we must use the full transformation $W = VU$ in Eq.~(\ref{W_app}). 
The procedure is applied sequentially, first dealing with the unitary part and then with the dissipative part. 
In this section we will focus only on the main results and leave the details of the calculations to Appendices~\ref{sec:gen_uni} and \ref{sec:gen_diss}. 

Once the functions $r_i(t)$ and $g_i(t)$ in Eqs.~(\ref{V_app}) and (\ref{U_app}) are properly adjusted, one finds the following surprisingly simple result for the rotating frame Liouvillian: 
\begin{equation}\label{Lt_QHO}
\tilde{\mathcal{L}} = \bar{\Lambda} \mathcal{H}_0  + \bar{\gamma} \mathcal{D}_1,
\end{equation}
where  $\Lambda(t) = \omega_t + 2 i \lambda_t r_2(t)$.
The variable $r_2(t)$ (which is part of the rotating frame transformation in Eq.~(\ref{U_app})), turns out to play a special role, being the time-periodic solution of the Riccati equation
\begin{equation}\label{r2}
\dot{r}_2 + 2 i \omega_t r_2 - 2 \lambda_t r_2^2 + \frac{\lambda_t^*}{2} = 0,
\end{equation}
which is the only non-linear equation in the problem.
The result in Eq.~(\ref{Lt_QHO}) is noteworthy. It shows that, in the generalized rotating frame, the system always evolves as a simple harmonic oscillator coupled to a zero-temperature bath with  damping rate $\bar{\gamma}$. 
We also note that in general, $\bar{\Lambda}$ may be complex, which is the source of potential instabilities, as explained below.

Next, applying Eq.~(\ref{LF}) we obtain for the Floquet Liouvillian (see Appendix~\ref{sec:gen_floquet}):
\begin{IEEEeqnarray}{rCl}\label{LF_QHO}
\mathcal{L}_F(t) &=& \omega_F(t) \mathcal{H}_0 + \frac{\lambda_F(t)}{2} \mathcal{H}_1 + \frac{\lambda_F'(t)}{2} \mathcal{H}_2 + \bar{\gamma} (N_F(t)+1) \mathcal{D}_1
\nonumber \\[0.2cm]
&& + \bar{\gamma} N_F(t) \mathcal{D}_2 - \bar{\gamma} M_F(t) \mathcal{D}_3 - \bar{\gamma} M_F'(t) \mathcal{D}_4.
\end{IEEEeqnarray}
This therefore has the same form of the original Liouvillian~(\ref{L}), but with modified parameters $\omega_F, \lambda_F, \lambda_F', N_F, M_F$ and $M_F'$, whose explicit forms are given below in Eqs.~(\ref{omegaF})-(\ref{MFp}). 
We note also that, in general, $\lambda_F' \neq \lambda_F^*$ and $M_F' \neq M_F^*$. 
However, this does not lead to unphysical results, as explained in Appendix~\ref{sec:gen_floquet}.
The steady-state of $\mathcal{L}_F(t)$ is also a squeezed thermal state, with covariances given by Eqs.~(\ref{ada2})-(\ref{adad2}). 
Thus, as in the purely dissipative case, knowing the Floquet Liouvillian immediately allows us to know the state in the limit cycle. 
We also call attention to the fact that the damping rate that appears in Eq.~(\ref{LF_QHO}) is $\bar{\gamma}$, which implies  that the steady-state will be  unique irrespective of how small $\gamma_t$, unless $\gamma_t = 0$ at all times.

When $\eta := \omega_t - \lambda_t$ is time-independent, we recover the more common mechanical situation of a harmonic oscillator 
$H = (p^2 + \Omega_t^2 q^2)/2$, 
subject to a time-periodic frequency $\Omega_t = \omega_t^2 - \lambda_t^2$. 
In this case we  may define a new variable $\xi(t)$ such that $r_2(t) = \frac{i}{2} + \xi^2/(i + i \xi^2 + \xi \dot{\xi}/\eta)$. 
Then Eq.~(\ref{r2}) implies that $\xi$ will satisfy the famous Ermakov-Pinney equation
\begin{equation}\label{Pinney}
\ddot{\xi} + \Omega_t^2 \xi = \frac{\eta^2}{\xi^3},
\end{equation}
which is exactly the same as in the unitary problem.
This equation always has a time-periodic solution, but it may either be real or such that $\xi^2$ is purely imaginary~\cite{Jeremie2010}. 
The former case corresponds to a stable evolution whereas the latter is unitarily unstable (that is, it would be unstable in the absence of dissipation). 
For a purely unitary evolution, these two regimes can be differentiated by the value of $\bar{\Lambda}$ appearing in Eq.~(\ref{Lt_QHO}), which is real in the unitarily stable phase and complex otherwise. 
These instabilities refer to the fact that depending on the drive, the trajectory of the harmonic oscillator may grow unboundedly, a problem which also occurs in the case of classical harmonic oscillators.

To know if a unitarily unstable solution will be stabilized by the presence of dissipation, we must look into the eigenvalues of $\tilde{\mathcal{L}}$ in Eq.~(\ref{Lt_QHO}).
Due to its simplicity, its eigenvalues can actually be found analytically and 
 read  $-\bar{\gamma} n/2 + 2i \bar{\Lambda} k$, where $n = 0,1,\ldots$ and $k \in [-\frac{n}{2},\frac{n}{2}]$, with $\Delta k = 1$.
Hence, we find that the condition for the system to converge to a stable  limit cycle is 
\begin{equation}\label{stab}
\bar{\gamma} > 2 | \text{Im}(\bar{\Lambda})|.
\end{equation}
This formula  provides a remarkably transparent method  for  determining the minimum amount of damping required to stabilize a cycle:
One must simply compare the output of the unitary evolution with the average damping.
This result holds for any type of protocol, hence generalizing and  simplifying the approach introduced in  Refs.~\cite{Insinga2018,Insinga2016}.

%
%
%
%

\section{\label{sec:carnot}Example: Carnot cycle}

Finally, to illustrate an application of the previous results, we  present the operation of a finite-time Carnot engine operating under continuous frequency modulations (c.f. Fig.~\ref{fig:drawing}(c)). 
We consider for simplicity the mechanical scenario where $\eta = \omega_t - \lambda_t$ is time-independent, so that the frequency protocol is completely specified by $\Omega_t = \omega_t^2 - \lambda_t^2$ (for concreteness we choose $\eta = \Omega_0$).
The order of the cycle is chosen as in Fig. 1(c):
\begin{itemize}
\item ($ab$) Hot isothermal expansion at $T_H$; 
\item ($bc$) Isentropic expansion;
\item  ($cd$) Cold isothermal compression at $T_C$; 
\item ($da$) Isentropic compression.
\end{itemize}
For the harmonic oscillator,  expansions (compressions) mean  decreasing (increasing) the frequency $\Omega$. 
All four strokes of the cycle were taken to have the same duration of $\mathcal{T}/4$. 

Before we turn to the finite time operations of the engine, it is necessary to review some properties of the quasi-static cycle. 
During an isothermal stroke, the energy at any time will be given by 
\begin{equation}
\langle H \rangle_t = \omega_t (N_t + \nicefrac{1}{2}) + \frac{\lambda_t M_t + \lambda_t^* M_t^*}{2} = \frac{\Omega_t}{2} \coth\bigg(\frac{\Omega_t}{2T}\bigg). 
\end{equation}
If $T \gg \Omega_t$  we get  the classical result $\langle H \rangle_t \simeq T$.
The classical harmonic oscillator therefore behaves like an ideal gas, in the sense that the energy during the isothermal stroke is constant. 
Conversely, for the quantum oscillator,  the energy depends on the frequency. 
In the isentropic strokes, on the other hand, the evolution is purely unitary so that the quasi-static energy is obtained from the adiabatic theorem and reads 
\begin{equation}
\langle H \rangle_t = \frac{\Omega_t}{\Omega_{t_0}}  \langle H \rangle_{t_0},
\end{equation}
where $t_0$ was the initial time of the unitary stroke. 

From these results we may then write down the energy of the system at the end of each quasi-static stroke: 
\begin{IEEEeqnarray}{rCl}
\langle H \rangle_a &=& \frac{\Omega_a}{\Omega_d} \langle H \rangle_d,
\\[0.2cm]
\langle H \rangle_b &=& \frac{\Omega_b}{2} \coth\bigg(\frac{\Omega_b}{2T_H}\bigg),	
\\[0.2cm]
\langle H \rangle_c &=&  \frac{\Omega_c}{\Omega_b} \langle H \rangle_b,
\\[0.2cm]
\langle H \rangle_d &=& \frac{\Omega_d}{2} \coth\bigg(\frac{\Omega_d}{2T_C}\bigg).	
\nonumber
\end{IEEEeqnarray}
We therefore see  that depending on the type of frequency protocol being used, the cycle may not have a \emph{reversible} quasi-static limit. 
The reason is that, if by the end of the isentropic strokes ($c$ and $a$) the energy $\langle H \rangle_c$ and $\langle H \rangle_a$ are not the same as the thermal equilibrium energies with the hot and cold baths respectively, then an inevitable dissipation will take place, even in the quasi-static limit. 
The condition for the existence of a reversible quasi-static limit is therefore obtained by imposing that
\[
\langle H \rangle_a = \frac{\Omega_a}{2}\coth\bigg(\frac{\Omega_a}{2T_H}\bigg),
\]
and
\[
\langle H \rangle_c = \frac{\Omega_c}{2} \coth\bigg(\frac{\Omega_c}{2T_C}\bigg).
\]
This therefore implies the constraints \cite{Sekimoto2000,Lekscha2016a}: 
\begin{equation}\label{Carnot_constraint}
\frac{T_C}{T_H} = \frac{\Omega_c}{\Omega_b} = \frac{\Omega_d}{\Omega_a},
\end{equation}
which we shall refer to the conditions for \emph{quasi-static reversibility}.

We now turn to the finite time operation.
With Eq.~(\ref{Carnot_constraint})  in mind, we choose for our cycle the frequency modulation
\begin{equation}\label{Omega_choice}
\Omega(t) = \Delta + \delta_t \cos^3(2\pi t/\mathcal{T}),
\end{equation}
where $\Delta = 1$ is a constant setting the overall energy scale. 
Moreover, $\delta_t$ is chosen so as to satisfy Eq.~(\ref{Carnot_constraint}), which implies
\begin{equation}
\delta_t = \begin{cases}
\delta 	& 	\text{ for } 0 < t < \mathcal{T}/4, 	\\[0.2cm]
\frac{\Delta \delta}{\Delta +\delta} 	& 	\text{ for } \mathcal{T}/4 < t <3 \mathcal{T}/4, \\[0.2cm]
\delta 	& 	\text{ for } 3\mathcal{T}/4 < t < \mathcal{T}.
\end{cases}
\end{equation}
In the results to be presented below, we have chosen for simplicity $\delta = 0.85$.
The choice~(\ref{Omega_choice}) for $\Omega(t)$ leads to a smooth function (only the third derivative is discontinuous), while still preserving the spirit of the Carnot cycle of having two expansion strokes followed by two compressions (see Fig~1(c)).

The damping rate was then chosen as 
\begin{equation}
\gamma_t = \begin{cases}
\gamma_0 	& 	\text{ for } 0 < t < \mathcal{T}/4, 	\\[0.2cm]
0 	& 	\text{ for } \mathcal{T}/4 < t < \mathcal{T}/2, \\[0.2cm]
\gamma_0 	& 	\text{ for } \mathcal{T}/2 < t < 3\mathcal{T}/4, 	\\[0.2cm]
0 	& 	\text{ for } 3\mathcal{T}/4 < t < \mathcal{T}.
\end{cases}
\end{equation}
This is illustrated in Fig.~1(c): the red and blue rectangles represent the two isothermals for which $\gamma_t = \gamma_0$ and the green rectangles represent the isentropic for which $\gamma_t = 0$. 
Finally, the Lindblad parameters $N_t$ and $M_t$ are chosen according to Eq.~(\ref{isothermal}):
\begin{equation}\label{isothermal}
N_t+ \nicefrac{1}{2} = \frac{\omega_t}{2\Omega_t} \coth\bigg(\frac{\Omega_t}{2T_t}\bigg),
\quad
M_t =  -\frac{\lambda_t}{2\Omega_t} \coth\bigg(\frac{\Omega_t}{2T_t}\bigg),
\end{equation}
where $T_t$ is given by 
\begin{equation}
T_t = \begin{cases}
T_H 	& 	\text{ for } 0 < t < \mathcal{T}/4, 	\\[0.2cm]
- 	& 	\text{ for } \mathcal{T}/4 < t < \mathcal{T}/2, \\[0.2cm]
T_C 	& 	\text{ for } \mathcal{T}/2 < t < 3\mathcal{T}/4, 	\\[0.2cm]
- 	& 	\text{ for } 3\mathcal{T}/4 < t < \mathcal{T}.
\end{cases}
\end{equation}
Here the notation ``-'' means that the value of the temperature in the isentropic strokes is irrelevant since $\gamma_t =0$. 
The hot temperature was chosen as $T_H = 1$ (in units of $\Delta = 1$). 
Then from Eq.~(\ref{Carnot_constraint}) one finds that the cold bath must have a temperature 
\begin{equation}
T_C = \frac{\Delta}{\delta + \Delta} T_H \simeq 0.54.
\end{equation}

\begin{figure}[!t]
\centering
\includegraphics[width=0.15\textwidth]{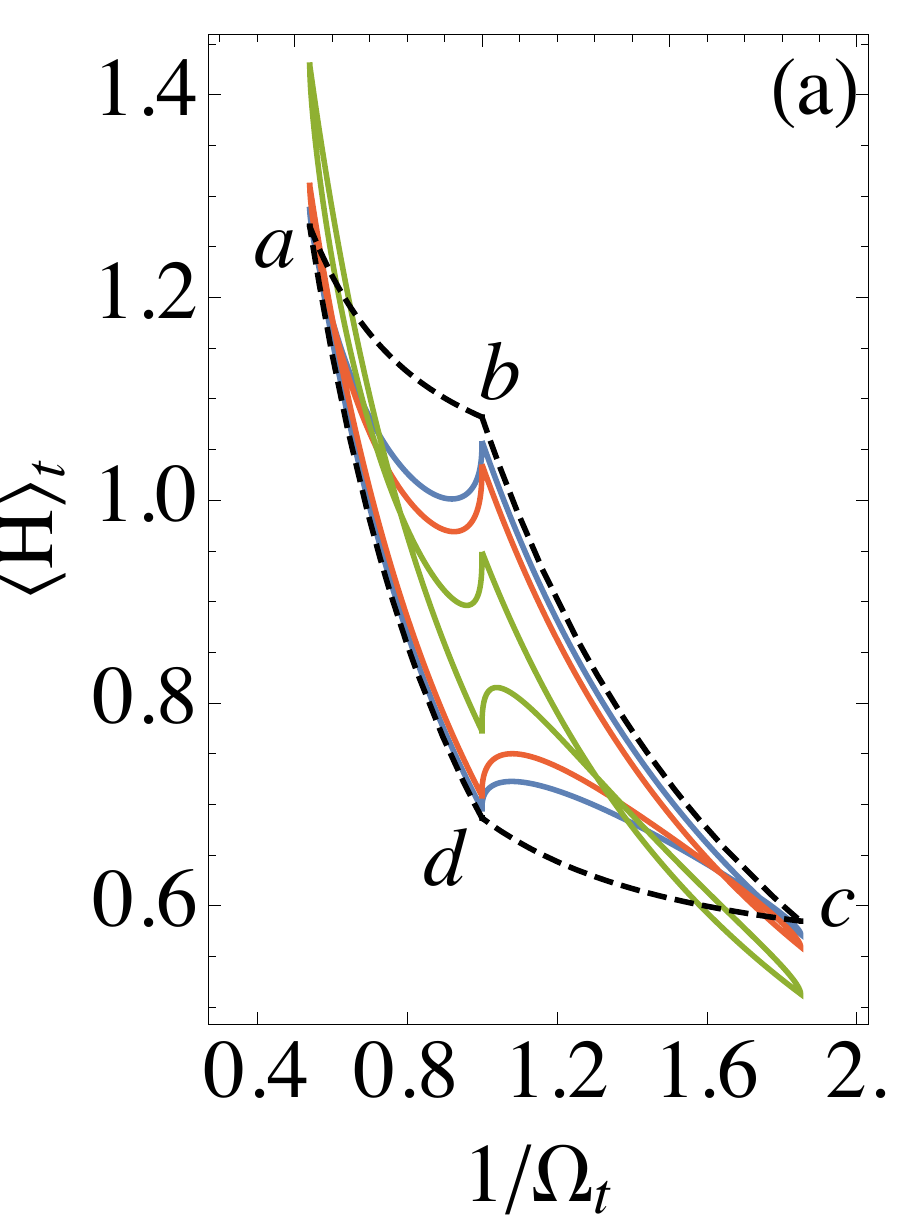}
\includegraphics[width=0.15\textwidth]{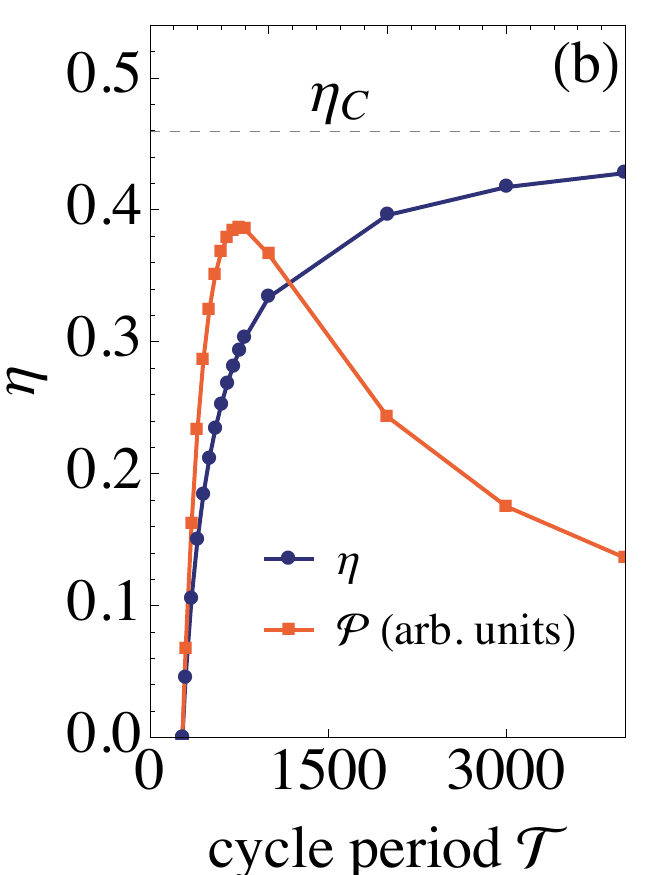}
\includegraphics[width=0.15\textwidth]{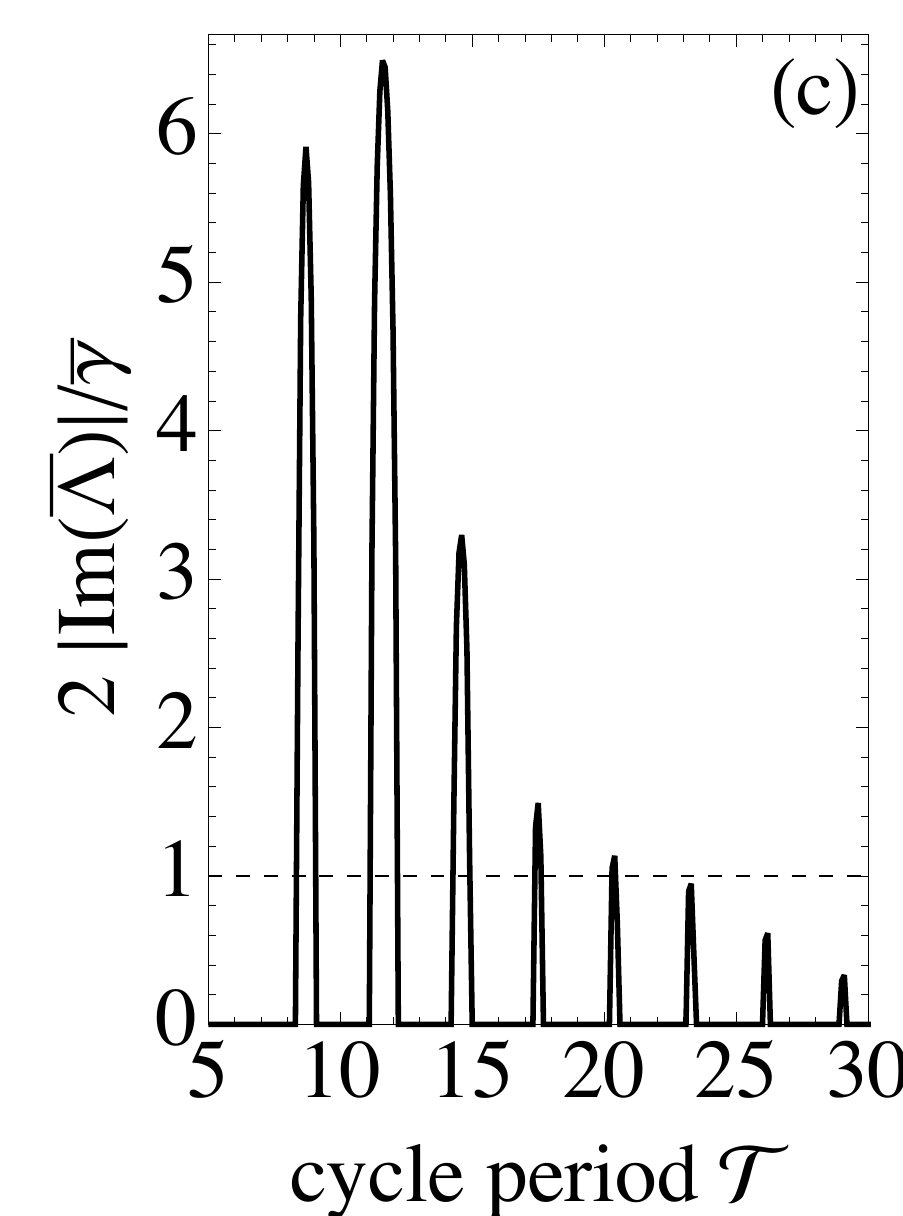}
\caption{\label{fig:carnot}
Operation of a finite-time Carnot engine.
The frequency modulation was chosen to be 
  $\Omega(t) = \Delta + \delta_t \cos^3(2\pi t/\mathcal{T})$ (Fig.~\ref{fig:drawing}(c)) and all  quantities are given in units of $\Delta = 1$. 
Quasi-static reversibility \cite{Sekimoto2000,Lekscha2016a} requires that we choose $\delta_t = \delta$ in $(ab,da)$ and $\delta_t = \Delta \delta/(\Delta +\delta)$ in $(bc,cd)$.
Other parameters are  $\delta = 0.85$, $T_H = 1$, $T_C = \Delta T_H/(\Delta + \delta)$ and $\gamma = 0.03$ in the isothermal strokes (and zero otherwise). 
(a) Stable cycles in a $\langle H \rangle_t$~vs.~$1/\Omega_t$ diagram for  periods $\mathcal{T} = 1000$, 700 and 300 (from outermost to innermost). 
The black dashed lines represent the corresponding quasi-static cycle.
(b) Efficiency $\eta = -  \mathcal{W} / Q_H $ (blue circles) and output power $\mathcal{P} = - \mathcal{W} /\mathcal{T}$ (red squares; arbitrary units)  as a function of the period $\mathcal{T}$. 
The uppermost dashed line corresponds to the classical Carnot efficiency $\eta_C = 1- T_C/T_H$.
Maximum power output is attained at $\mathcal{T}\sim 700$. 
(c) Stability analysis using Eq.~(\ref{stab}), obtained by plotting $2|\text{Im}(\bar{\Lambda})|/\bar{\gamma}$~vs.~$\mathcal{T}$.
When this quantity is larger than unity (horizontal dashed line) the cycle becomes unstable.
}
\end{figure}

Examples of finite time cycles are shown in Fig.~\ref{fig:carnot}(a), in which the gradual convergence towards the quasi-static limit can be clearly observed. 
In Fig.~\ref{fig:carnot}(b) we present the efficiency and the output power. 
The analysis of the quasi-static work, heat and efficiency is presented in Appendix~\ref{sec:quasi}. 
As expected, when the cycle duration $\mathcal{T}$ becomes large the efficiency tends to the Carnot efficiency and the output power tends to zero. 
Maximum power output is attained at $\mathcal{T}\sim 700$. 
Finally, we consider the stability of the cycle in Fig.~\ref{fig:carnot}(c), by studying  Eq.~(\ref{r2}) and the stability criteria~(\ref{stab}).
These results show that the regions of instability appear in the form of pulses, signifying a type of resonant behavior. 
The interesting aspect of these results is that it allows one to devise the necessary amount of dissipation required to create a stable cycle.

\section{Concluding remarks}

We have shown how Lindblad-Floquet theory can be used to provide a unified description of finite time quantum heat engines, answering both the question of convergence to a limit cycle and giving the state of the system once the limit-cycle has been reached. 
As an example, we applied it to an open harmonic oscillator, which is the most widely used working fluid used quantum heat engines.  
As this example has shown,  casting the problem in this framework allows for a deeper understanding of the dynamical features involved in the cycle's operation. 
Of course, finding the  Floquet Liouvillian in practice may be a difficult task. 
The example we have given is one for which this task can be done analytically,  something which is in principle true whenever the algebra of the super-operators is closed. 
Otherwise, it must be found perturbatively.

We believe that the results presented here could serve as a platform for advancing our understanding of quantum heat engines. 
In particular, two immediate applications stand out. 
First, this theory provides the ideal framework for designing shortcut to adiabaticity protocols operating in the finite-time scenario. 
Secondly, it encompasses naturally the use of time-dependent squeezing modulations, which could be used to reach higher efficiency at maximum power. 
Other physical implementations, such as qubit systems or multiple bosonic modes, can also be studied using the same methods.

\emph{Acknowledgments -}
The authors acknowledge the USP-COFECUB project number Uc Ph 167-17. 
GTL would like to acknowledge the S\~ao Paulo Research Foundation, under grant number 2016/ 08721-7.
The authors acknowledge fruitful discussions with Mauro Paternostro, Obinna Abah, Oussama Houhou and Jader Santos. 

\appendix
\section{\label{sec:relation}Relation between the bosonic and mechanical pictures}

Let us assume that the parameter $\lambda_t$ appearing in the Hamiltonian~(\ref{H_QHO}) is real and let us rewrite $H$ in terms of position and momentum. 
To do so for time-dependent Hamiltonians requires some care since naive transformations would lead to time-dependent quadrature operators. 
In order to keep on having time-independent operators, one must then introduce a transformation of the form 
\begin{equation}
q = \frac{1}{\sqrt{2\eta}} (a^\dagger + a),
\qquad 
p = i \sqrt{\frac{\eta}{2}} (a^\dagger - a).
\end{equation}
where $\eta$ is an arbitrary frequency scale setting the units of $q$ and $p$. 
In terms of these new variables, the bosonic Hamiltonian~(\ref{H_QHO}) becomes
\begin{equation}\label{H2_QHO}
H = \frac{(\omega_t-\lambda_t)}{2\eta} p^2 + \frac{\eta}{2} (\omega_t+\lambda_t) q^2.
\end{equation}
Thus, we see that in general the Hamiltonian~(\ref{H_QHO}) corresponds to a mechanical oscillator where both the frequency and the mass are time-dependent. 

We also see that the more usual problem of a time-independent mass occurs when   $\omega_t - \lambda_t$ is time-independent.
In this case one may choose, without loss of generality,  $\eta = \omega_t - \lambda_t$, leading to 
\begin{equation}\label{H3_QHO}
H = \frac{p^2}{2} + \frac{\Omega_t^2}{2} q^2,
\end{equation}
where
\begin{equation}\label{Omega_def}
\Omega_t^2 = \omega_t^2 - \lambda_t^2 = \eta(\omega_t + \lambda_t).
\end{equation}
In this case one therefore recovers precisely the setup of the Ermakov-Lewis theory.

\section{\label{sec:gen_uni}Unitary part}

In this appendix we consider the problem of tackling the full time-dependent  Liouvillian~(\ref{L}). 
Using the transformation~(\ref{W_app}) in 
Eq.~(\ref{Lt}) leads to
\begin{equation}\label{Lt2}
\tilde{\mathcal{L}} = \frac{\ud V}{\ud t} V^{-1} + V \bigg\{ \frac{\ud U}{\ud t} U^{-1}  + U \mathcal{L}_t U^{-1} \bigg\} V^{-1}.
\end{equation}
We will begin by dealing with the terms  inside curly brackets, corresponding to the unitary evolution.

Carrying out the BCH expansions, as in Sec.~\ref{ssec:purely}, we find
\begin{equation}\label{Lt3}
\frac{\ud U}{\ud t} U^{-1}  + U \mathcal{L}_t U^{-1}  = B_0(t) \mathcal{H}_0 + B_1(t) \mathcal{H}_1 + B_2(t) \mathcal{H}_2 + \tilde{\mathcal{D}}_t,
\end{equation}
where $\tilde{\mathcal{D}}_t = U \mathcal{D}_t U^{-1}$ (to be dealt with in Appendix~\ref{sec:gen_diss}) and 
\begin{IEEEeqnarray}{rCl}
\label{B2}
B_2(t) &=& e^{-2i r_0} \bigg[ \dot{r}_2 + 2 i \omega_t r_2 - 2 \lambda_t r_2^2 + \frac{\lambda_t^*}{2} \bigg],		\\[0.2cm]
\label{B1}
B_1(t) &=& -4 r_1^2 B_2(t) + e^{2 i r_0} \bigg[ \dot{r}_1 - 2 i \omega_t r_1 + 4 \lambda_t r_1 r_2 + \frac{\lambda_t}{2} \bigg],	\\[0.2cm]
\label{B0}
B_0(t) &=& - 4 i r_1 e^{2 i r_0} B_2(t) + \dot{r}_0 + \omega_t + 2 i \lambda_t r_2 .
\end{IEEEeqnarray}
Next we adjust the functions $r_0$, $r_1$ and $r_2$ so as to make the unitary part of Eq.~(\ref{Lt3}) time-independent.
To accomplish this, we  choose $r_1(t)$ and $r_2(t)$ to be the time-periodic solutions of 
\begin{IEEEeqnarray}{rCl}
\label{uni_r2}
\dot{r}_2 + 2 i \omega r_2 - 2 \lambda_t r_2^2 + \frac{\lambda_t^*}{2} &=& 0 ,		\\[0.2cm]
\label{uni_r1}
\dot{r}_1 - 2 i \omega r_1 + 4 \lambda_t r_1 r_2 + \frac{\lambda_t}{2} &=& 0 ,
\end{IEEEeqnarray}
which then makes $B_1 = B_2 = 0$. 

Next let 
\begin{equation}\label{Lambda}
\Lambda(t) = \omega_t + 2 i \lambda_t r_2 (t).
\end{equation}
Then, to make  $B_0(t)$ in Eq.~(\ref{B0})  time-independent we  choose
\begin{equation}\label{uni_r0}
r_0(t)  = \int\limits_0^t \ud t' \bigg[ \bar{\Lambda} - \Lambda(t')\bigg],
\end{equation}
where, recall, the time-average is defined in Eq.~(\ref{time_average}).
With these choices  Eq.~(\ref{Lt3}) becomes 
\begin{equation}\label{Lt4}
\frac{\ud U}{\ud t} U^{-1}  + U \mathcal{L}_t U^{-1}  = \bar{\Lambda} \mathcal{H}_0 + \tilde{\mathcal{D}}_t.
\end{equation}
Consequently, Eq.~(\ref{Lt2}) reduces to
\begin{equation}\label{Lt5}
\tilde{\mathcal{L}} = \frac{\ud V}{\ud t} V^{-1} + V \bigg\{ \bar{\Lambda} \mathcal{H}_0 + \tilde{\mathcal{D}}_t \bigg\} V^{-1}.
\end{equation}
The next step is to now turn to the dissipative part and adjust $V(t)$ in order to make Eq.~(\ref{Lt5}) time-independent. 

But before doing so it is useful to anticipate certain facts about $r_1$ and $r_2$.
Eq.~(\ref{uni_r2}) admits two types of solutions, representing unitarily stable (US) and unitarily unstable (UU) dynamics (by ``unitarily'' we refer to stability in the absence of dissipation). 
Let us define the variable
\begin{equation}\label{sigma}
\sigma = \begin{cases} 
1& \text{ Unitarily Stable (US)}	\\[0.2cm]
i & \text{ Unitarily Unstable (UU)}
\end{cases}
\end{equation}
Then, the properties of the  two phases are most readily distinguished by means of the following auxiliary variables: 
\begin{IEEEeqnarray}{rCl}
\label{J}
J &=& 1 + 8 r_1 r_2 	,
\\[0.2cm]
\label{z}
z &=& 1 + 4 r_1 r_2 = \frac{1+J}{2},
\\[0.2cm]
\label{r1p}
r_1' &=& r_2 (1+ 4 r_1 r_2) = r_2 z,
\end{IEEEeqnarray}
which  are introduced to make the results that follow more self-contained.
As will be discussed in Appendix~\ref{sec:gen_EL}, it turns out that 
\begin{equation}
\label{r1p_res}
r_1'=  \sigma^2 r_1^* ,
\end{equation}
and
\begin{equation}\label{J_res}
J = \sigma j ,\qquad j \in \mathbb{R}.
\end{equation}
Using also that $J^2 = 1 + 16 r_1 r_1' = 1+ 16 \sigma^2 |r_1|^2$ we find that 
\begin{equation}
j^2 = 16 |r_1|^2 + \sigma^2.
\end{equation}
Moreover, combining these results we find
\begin{equation}\label{r1_z}
4 |r_1|^2 = \sigma^2 \left(\frac{J^2-1}{4} \right)= \sigma^2 z (z-1).
\end{equation}
Finally, it is worth mentioning that 
\begin{equation}
z^* 
= \begin{cases} 
z 	& 	\qquad \sigma = 1			\\[0.2cm]
1-z 	& 	\qquad \sigma = -i 		
\end{cases}.
\label{z_cases}
\end{equation}

We can also use the above results to express $r_2$ in terms of $r_1$, $z$ and $j$ in various ways:
\begin{equation}
r_2 = \frac{\sigma^2 r_1^*}{z}= \frac{\sigma j-1}{8 r_1} = \frac{2 \sigma^2 r_1^*}{\sigma j + 1}.
\end{equation}
In particular, it then follows that in the UU phase ($\sigma = -i$) 
\begin{equation}
4|r_2|^2 = 1.
\end{equation}
so $r_2(t)$ evolves in time as a pure phase.

\section{\label{sec:gen_diss}Dissipative part}

We now return to Eq.~(\ref{Lt5}) and adjust the functions $g_i$ in order to eliminate the remaining time-dependence.
To do so we first need to compute $\tilde{\mathcal{D}}_t = U \mathcal{D}_t U^{-1}$. 
Using again the BCH expansions we find that $\tilde{\mathcal{D}}_t$ has the same structure as $\mathcal{D}_t$ in Eq.~(\ref{Dt}), but with modified parameters
\begin{IEEEeqnarray}{rCl}
\tilde{\mathcal{D}}_t 
&=&  \gamma_t (\tilde{N}_t +1) \mathcal{D}_1 + \gamma_t \tilde{N}_t \mathcal{D}_2  - \gamma_t \tilde{M}_t \mathcal{D}_3 - \gamma_t \tilde{M}_t' \mathcal{D}_4,
\end{IEEEeqnarray}
where
\begin{IEEEeqnarray}{rCl}
\label{Ntilde}
\tilde{N}_t +\nicefrac{1}{2}&=&J( N_t+\nicefrac{1}{2}) + 2 i M_t r_1 - 2 i M_t^* r_1',	\\[0.2cm]
\label{Mtilde}
\tilde{M}_t &=& \bigg[ M_t - 4 i r_2 (N_t + \nicefrac{1}{2}) - 4 M_t^* r_2^2 \bigg] e^{-2i r_0},	\\[0.2cm]
\label{Mptilde}
\tilde{M}_t' &=& \bigg[ M_t^* z^2 + 4 i r_1 z (N_t + \nicefrac{1}{2}) - 4 M_t r_1^2 \bigg] e^{2i r_0}.
\end{IEEEeqnarray}
In these formulas, we assume that we have already solved for the $r_i$, so that these correspond simply to new time-periodic parameters. 
In general, however, $\tilde{M}_t' \neq \tilde{M}_t^*$ and $N_t$ may now be complex. 
Below in this subsection we will show how that can be amended. 

We now see from this result that Eq.~(\ref{Lt5}) falls under the same category of the problem treated in Sec.~\ref{ssec:purely}. 
Carrying out all expansions we find
\begin{equation}\label{Lt6}
\tilde{\mathcal{L}} = \bar{\Lambda} \mathcal{H}_0 + C_1(t) \mathcal{D}_1 + C_2(t) \mathcal{D}_2 + C_3(t) \mathcal{D}_3 + C_4(t) \mathcal{D}_4,
\end{equation}
where now 
\begin{IEEEeqnarray}{rCl}
\label{C2}
C_2(t) &=& e^{-g_1} \bigg[ \dot{g}_2 - \gamma_t + \gamma_t e^{g_2} (\tilde{N}_t+1)\bigg],	\\[0.2cm]
\label{C1}
C_1(t) &=& \dot{g}_1 - \dot{g}_2 + \gamma_t + C_2 ,		\\[0.2cm]
\label{C3}
C_3(t) &=& e^{g_2 - g_1} \bigg[ \dot{g}_3 + (\gamma_t + 2 i \bar{\Lambda}) g_3 - \gamma_t \tilde{M}_t \bigg] ,	\\[0.2cm]
\label{C4} 
C_4(t) &=& e^{g_2 - g_1} \bigg[ \dot{g}_4 + (\gamma_t - 2 i \bar{\Lambda}) g_4 - \gamma_t \tilde{M}_t' \bigg].
\end{IEEEeqnarray}
The only difference with respect to Eqs.~(\ref{diss_C2})-(\ref{diss_C4}) is the appearance of $\bar{\Lambda}$ and the fact that the physical parameters $(N_t,M_t,M_t^*)$ are now replaced by $(\tilde{N}_t, \tilde{M}_t, \tilde{M}_t')$. 
Proceeding as before, we then obtain a time-independent $\tilde{\mathcal{L}}$ by setting 
\begin{IEEEeqnarray}{rCl}
\label{full_g2}
\dot{g}_2 - \gamma_t + \gamma_t e^{g_2} (\tilde{N}_t+1) &=& 0 ,	\\[0.2cm]
\label{full_g3}
\dot{g}_3 + (\gamma_t + 2 i \bar{\Lambda}) g_3 &=& \gamma_t \tilde{M}_t, \\[0.2cm]
\label{full_g4}
\dot{g}_4 + (\gamma_t - 2 i \bar{\Lambda}) g_4 &=&  \gamma_t \tilde{M}_t' .
\end{IEEEeqnarray}
and 
\begin{equation}\label{full_g1}
g_1(t) = g_2(t) + \int\limits_0^t \ud t' \bigg[\bar{\gamma} - \gamma(t')\bigg]	.
\end{equation}
We also define $G_2(t)$ exactly as in Eq.~(\ref{G2_def}), which then gives 
\begin{equation}\label{full_G2}
\dot{G_2} + \gamma_t G_2 = \gamma_t (\tilde{N}_t+  \nicefrac{1}{2}).
\end{equation}
After setting all these functions, we then finally obtain 
\begin{equation}\label{Lt_final}
\tilde{\mathcal{L}} =\bar{\Lambda} \mathcal{H}_0 + \bar{\gamma} \mathcal{D}_1.
\end{equation}
which  is Eq.~(\ref{Lt_QHO}).

In the above formulation we generally have $g_4 \neq g_3^*$ and $G_2$ complex.
It is therefore convenient to use a new set of variables in which the physical meaning of these variables can be made clearer.
The variable $G_2$ can be left as is. But it is convenient to use Eqs.~(\ref{sigma})-(\ref{z_cases}) to rewrite Eq.~(\ref{Ntilde}) as 
\begin{IEEEeqnarray}{rCl}
\label{Ntilde2}
\tilde{N}_t +\nicefrac{1}{2}&=& \sigma \bigg\{ j(N_t+\nicefrac{1}{2}) + 2 i M_t r_1 \sigma^* - 2i M_t^* r_1^* \sigma\bigg\}.
\end{IEEEeqnarray}
The quantity inside brackets in Eq.~(\ref{Ntilde2}) is now real by construction, so that  $\tilde{N}_t + \nicefrac{1}{2} \propto \sigma$.
Consequently, the same will be true for $G_2$. 
As  will be seen below, $G_2$ always appears in products of the form $J G_2 \propto \sigma^2$, which will therefore be real. 

Next we turn to $g_3$ and $g_4$.
First we eliminate the dependence on $r_0$ by defining $\tilde{g}_3 = e^{2 i r_0} g_3$ and $\tilde{g}_4 = e^{-2 i r_0} g_4$. 
Because of Eq.~(\ref{uni_r0}) it then follows that Eqs.~(\ref{full_g3}) and (\ref{full_g4}) are simply replaced by 
\begin{IEEEeqnarray}{rCl}
\label{full_g3_3}
\dot{\tilde{g}}_3 + (\gamma_t + 2 i \Lambda_t) \tilde{g}_3 &=& \gamma_t 
\bigg[ M_t - 4 i r_2 (N_t + \nicefrac{1}{2}) - 4 M_t^* r_2^2 \bigg],
\IEEEeqnarraynumspace \\[0.2cm]
\label{full_g4_3}
\dot{\tilde{g}}_4 + (\gamma_t - 2 i\Lambda_t) \tilde{g}_4 &=&  \gamma_t
\bigg[ M_t^* z^2 + 4 i r_1 z (N_t + \nicefrac{1}{2}) - 4 M_t r_1^2 \bigg].
\IEEEeqnarraynumspace
\end{IEEEeqnarray}
That is, compared to Eqs.~(\ref{full_g3}) and (\ref{full_g4}), $\bar{\Lambda}$ is replaced by $\Lambda$ and the factors of $e^{\pm 2i r_0}$ are eliminated from Eqs.~(\ref{Mtilde}) and (\ref{Mptilde}). 

Next we change variables to 
\begin{equation}\label{G3G4}
G_3 = z  \tilde{g}_3, \qquad
G_4 = \frac{ \tilde{g}_4}{z}.
\end{equation}
where, recall, $z = 1 + 4 r_1 r_2$ [Eq.~(\ref{z})]. 
We then get 
\begin{widetext}
\begin{IEEEeqnarray}{rCl}
\label{G3_2}
\dot{G}_3 + (\gamma_t + 2 i \omega_t- \nu_t) G_3&=& \gamma_t\bigg[M_t z - 4 i  \sigma^2 r_1^* (N_t+\nicefrac{1}{2}) - \frac{4 M_t^*(r_1^*)^2}{z} \bigg],
\IEEEeqnarraynumspace\\[0.2cm]
\dot{G}_4+ (\gamma_t - 2 i \omega_t + \nu_t) G_4 &=& \gamma_t\bigg[
M_t^* z +4 i  r_1 (N_t+\nicefrac{1}{2}) - \frac{4 M_t(r_1)^2}{z} 
\bigg].
\IEEEeqnarraynumspace
\end{IEEEeqnarray}
where
\begin{IEEEeqnarray}{rCl}
\nu_t &=& 4 \lambda_t r_2 + \frac{\dot{z}}{z} \\[0.2cm]
&=&2 \lambda_t r_2 - \frac{2 \lambda^* r_1}{z}
\\[0.2cm]
&=&  \frac{2\lambda_t \sigma^2 r_1^* - 2\lambda^* r_1}{z}.
\end{IEEEeqnarray}
In the US phase $\sigma = 1$ and $z^* =z $ so that $\nu^* = - \nu$. 
Consequently, we see that in this case $G_3^* = G_4$. 
In the UU phase, on the other hand, this is no longer true. 
However, in the UU phase a new symmetry appears. Namely, $i r_1 G_3$ and $i r_1^* G_4$  become real.
This can be seen by verifying that $i r_1 G_3$ satisfies a linear real equation, so that the solution must also be real. 
To summarize:
\begin{itemize}
\item In the Unitarily Stable (US) case ($\sigma = 1$) we have $G_3 = G_4^*$ and $G_2 \in \mathbb{R}$. 
\item In the Unitarily Unstable (UU) case ($\sigma = i$) we have $(r_1 G_3)^* = - (r_1 G_3)$,  $(r_1^* G_4)^* = - r_1^* G_4$ and $i G_2 \in \mathbb{R}$. 
\end{itemize}

\section{\label{sec:gen_floquet}Floquet Liouvillian}

The final step is to apply Eq.~(\ref{LF}) to find the Floquet Liouvillian. 
The  result is Eq.~(\ref{LF_QHO}) with 
\begin{IEEEeqnarray}{rCl}
\label{omegaF}
\omega_F &=& \bar{\Lambda} J, 	\\[0.2cm]
\lambda_F &=& 4 i \bar{\Lambda} r_1 ,		\\[0.2cm]
\lambda_F' &=& -4 i\bar{\Lambda} r_1'	,\\[0.2cm]
N_F +\nicefrac{1}{2}&=&J G_2 - \frac{2 i}{\bar{\gamma}} \bigg[ r_1  z   (\bar{\gamma} + 2 i \bar{\Lambda})  \tilde{g}_3 -r_2  (\bar{\gamma} - 2 i \bar{\Lambda})   \tilde{g}_4\bigg],
\IEEEeqnarraynumspace\\[0.2cm]
M_F &=& 4 i  r_1' G_2 + \frac{1}{\bar{\gamma}} \bigg[z^2 (\bar{\gamma} + 2 i \bar{\Lambda}) \tilde{g}_3 - 4 r_2^2  (\bar{\gamma} - 2 i \bar{\Lambda} ) \tilde{g}_4 \bigg], \IEEEeqnarraynumspace	\\[0.2cm]
M_F' &=& -4 i r_1 G_2 + \frac{1}{\bar{\gamma}} \bigg[ -4 r_1^2 (\bar{\gamma} + 2 i \bar{\Lambda}) \tilde{g}_3 + (\bar{\gamma} - 2i \bar{\Lambda}) \tilde{g}_4 \bigg] .
\label{MFp}\IEEEeqnarraynumspace
\end{IEEEeqnarray}
In these expressions, all quantities are to be evaluated at time $t$, which has been omitted for clarity.

The steady-state of the Floquet Liouvillian~(\ref{LF}) is unique and corresponds to a squeezed thermal state, which is completely characterized by the second moments
\begin{IEEEeqnarray}{rCl}
\label{ada}
\langle a^\dagger a \rangle +\nicefrac{1}{2}&=& 
\frac{(N_F + \nicefrac{1}{2}) (\bar{\gamma}^2 + 4 \omega_F^2) + M_F \lambda_F (2\omega_F + i \bar{\gamma}) + M_F' \lambda_F' (2 \omega_F - i \bar{\gamma})}{\bar{\gamma}^2 + 4 \omega_F^2 - 4 \lambda_F \lambda_F'},
\IEEEeqnarraynumspace
\\[0.2cm]
\langle a a \rangle &=&
\frac{M_F (\bar{\gamma}^2 - 2 i \bar{\gamma} \omega_F - 2 \lambda_F \lambda_F') - 2 M_F' \lambda_F'^2 - \lambda_F' (2 N_F+1) (2\omega_F + i \bar{\gamma})}{\bar{\gamma}^2 + 4 \omega_F^2 - 4 \lambda_F \lambda_F'},
\IEEEeqnarraynumspace\\[0.2cm]
\langle a^\dagger a^\dagger \rangle &=&
\frac{M_F' (\bar{\gamma}^2 + 2 i \bar{\gamma} \omega_F - 2 \lambda_F \lambda_F') - 2 M_F \lambda_F^2 - \lambda_F (2 N_F+1) (2\omega_F - i \bar{\gamma})}{\bar{\gamma}^2 + 4 \omega_F^2 - 4 \lambda_F \lambda_F'}.
\IEEEeqnarraynumspace
\label{adad}
\end{IEEEeqnarray}
\end{widetext}
Substituting the Floquet parameters Eqs.~(\ref{omegaF})-(\ref{MFp}) into Eqs.~(\ref{ada})-(\ref{adad}) we get
\begin{IEEEeqnarray}{rCl}
\label{ada2}
\langle a^\dagger a \rangle +\nicefrac{1}{2}&=& 
J G_2- 2 i r_1 z \tilde{g}_3 + 2 i r_2  \tilde{g}_4,
\IEEEeqnarraynumspace\\[0.2cm]
\langle a a \rangle &=&
4 i r_1' G_2  + z^2 \tilde{g}_3 - 4 r_2^2 \tilde{g}_4,
\IEEEeqnarraynumspace\\[0.2cm]
\langle a^\dagger a^\dagger \rangle &=&
-4 i r_1 G_2  - 4 r_1^2 \tilde{g}_3 + \tilde{g}_4.
\label{adad2}\IEEEeqnarraynumspace
\end{IEEEeqnarray}
In terms of the variables $G_3$ and $G_4$, defined in Eq.~(\ref{G3G4}), these simplify even further to
\begin{IEEEeqnarray}{rCl}
\label{ada3}
\langle a^\dagger a \rangle +\nicefrac{1}{2}&=& 
J G_2- 2 i (r_1 G_3 - r_1' G_4),
\IEEEeqnarraynumspace\\[0.2cm]
\label{aa3}
\langle a a \rangle &=&
4 i r_1' G_2  + z G_3 -  \frac{4r_1'^2}{z} G_4,
\IEEEeqnarraynumspace\\[0.2cm]
\langle a^\dagger a^\dagger \rangle &=&
-4 i r_1 G_2  -  \frac{4r_1^2}{z} G_3 + z G_4.
\label{adad3}\IEEEeqnarraynumspace
\end{IEEEeqnarray}

It follows from this result, together with our previous discussion about $G_2$, $G_3$ and $G_4$,  that in both the US and UU cases we will have $\langle a^\dagger a \rangle$ real and $\langle a a\rangle = \langle a^\dagger a^\dagger \rangle^*$, as expected on physical grounds. 
This is a bit cumbersome to verify but can be done as follows. 
In the US phase $r_1 G_3 - r_1' G_4 = r_1 G_3 - r_1^* G_3^*$ which is purely imaginary, hence making~(\ref{ada3}) real. 
In the UU phase, on the other hand, $r_1 G_3$ and $r_1^* G_4$ will independently be purely imaginary, hence leading to the same conclusion. 
One may proceed similarly when comparing~(\ref{aa3}) and (\ref{adad3}). 
For instance, in the US phase $(z G_3 - 4 r_1^{*2} G_4/z)^* = z G_4 - 4 r_1^2 G_3/z$, hence making $\langle aa \rangle^* = \langle a^\dagger a^\dagger \rangle$. 
In the UU phase, on the other hand, one has to make use of Eq.~(\ref{r1_z}), which in this case is written as $4|r_1|^2 = z z^*$. 
Then since $G_3^* = - r_1 G_3/r_1^*$ it follows that $(z G_3)^* = - z^* r_1 G_3/r_1^* = - 4 r_1^2 G_3/z$.  
A similar calculation will hold for $\frac{4r_1*^2}{z} G_4$ so that, once again, we will have $\langle aa \rangle^* = \langle a^\dagger a^\dagger \rangle$.

\section{\label{sec:gen_EL}Stability of the Ermakov-Lewis theory}

Let us analyze the connection between our results and the Ermakov-Lewis theory describing the unitary dynamics of a harmonic oscillator subject to a time-dependent frequency. 
As discussed in Sec.~\ref{sec:relation}, this connection is established when $\eta = \omega_t - \lambda_t$ is time-independent, in which  case we  work instead with $\Omega_t$ defined in Eq.~(\ref{Omega_def}). 
If we now let let
\begin{IEEEeqnarray}{rCl}
\label{r1_pinney}
r_1(t) &=& \frac{\xi \dot{\xi}}{4 \eta} - \frac{i}{8} \bigg[ \frac{1}{\xi^2} - \xi^2 + \frac{\dot{\xi}^2}{\eta^2}\bigg],
\\[0.2cm]
\label{r2_pinney}
r_2(t) &=& \frac{i}{2} + \frac{\xi^2}{i (1 + \xi^2) + \xi \dot{\xi}/\eta}.
\end{IEEEeqnarray}
then one may verify that Eq.~(\ref{uni_r2}) will be satisfied provided $\xi$ is a time-periodic solution of the Pinney equation~(\ref{Pinney})
This therefore serves to show that our calculations reproduce the Ermakov-Lewis theory as a particular case.

Eq.~(\ref{Pinney}) always admits a time-periodic solution. 
However, this solution may be either real or such that $\xi^2$ is purely imaginary. 
The former corresponds to unitarily stable solutions whereas the latter are unitarily unstable  (that is, they would be unstable in the absence of dissipation). 
This can now be used to demonstrate the facts stated below Eqs.~(\ref{J}) and (\ref{r1p}).
The quantities $J$ and $r_1'$ may be rewritten as
\begin{IEEEeqnarray}{rCl}
J &=& 1 + 8 r_1 r_2  = \frac{1}{2} \bigg( \frac{1}{\xi^2} + \xi^2 + \frac{\dot{\xi}^2}{\eta^2} \bigg)	,	\\[0.2cm]
r_1' &=& r_2 (1+ 4 r_1 r_2) = \frac{\xi \dot{\xi}}{4 \eta} + \frac{i}{8} \bigg[ \frac{1}{\xi^2} - \xi^2 + \frac{\dot{\xi}^2}{\eta^2}\bigg].
\end{IEEEeqnarray}
From the properties of $\xi$ in the two phases it then readily follows that $J^* = J$ in the US phase and $J^* = - J$ in the UU phase [Eq.~(\ref{J_res})].
Similarly, it follows that $r_1' = r_1^*$ in one case and $r_1' = - r_1^*$ in the other [Eq.~(\ref{r1p_res})].

\section{\label{sec:quasi}Heat, work and efficiency in the quasi-static case}

In this appendix we discuss the quasi-static properties of the Carnot cycle. 
We take for the work rate the usual definition 
\begin{equation}
\frac{\ud \mathcal{W}}{\ud t} = \bigg\langle \frac{\partial H}{\partial t} \bigg\rangle = \dot{\omega}_t (\langle a^\dagger a \rangle_t + \nicefrac{1}{2}) + \frac{\dot{\lambda}_t}{2} (\langle a a \rangle_t + \langle a^\dagger a^\dagger \rangle_t) .
\end{equation}
In the isentropic strokes no heat flows to the environment so that the total work performed becomes simply 
\begin{IEEEeqnarray}{rCl}
\mathcal{W}_{bc} &=& \langle H \rangle_c - \langle H \rangle_b 
\\[0.2cm]
&=& \frac{\Omega_c}{2} \coth\bigg(\frac{\Omega_c}{2T_C}\bigg) - 
\frac{\Omega_b}{2} \coth\bigg(\frac{\Omega_b}{2T_H}\bigg),
\\[0.2cm]
\mathcal{W}_{da} &=& \langle H \rangle_a - \langle H \rangle_d \\[0.2cm]
&=& \frac{\Omega_a}{2} \coth\bigg(\frac{\Omega_a}{2T_H}\bigg) - 
\frac{\Omega_d}{2} \coth\bigg(\frac{\Omega_d}{2T_C}\bigg).
\end{IEEEeqnarray}

As for the isothermal strokes, we have 
\begin{IEEEeqnarray}{rCl}
\langle a^\dagger a \rangle_t + \nicefrac{1}{2} &=& N_t + \nicefrac{1}{2} = \frac{\omega_t}{2\Omega_t} \coth\bigg(\frac{\Omega_t}{2T}\bigg) ,
\\[0.2cm]
\langle a a \rangle_t &=& M_t =- \frac{\lambda_t}{2\Omega_t} \coth\bigg(\frac{\Omega_t}{2T}\bigg) ,
\end{IEEEeqnarray}
where $T$ means either $T_H$ or $T_C$.
The work rate then becomes
\begin{equation}
\frac{\ud \mathcal{W}}{\ud t}  = \frac{\dot{\omega}_t \omega_t - \dot{\lambda}_t\lambda_t}{2 \Omega_t} \coth\bigg(\frac{\Omega_t}{2T}\bigg) .
\end{equation}
But this may be written as 
\begin{equation}
\frac{\ud \mathcal{W}}{\ud t}  =  T \frac{\ud }{\ud t} \ln \bigg[ \sinh\bigg( \frac{\Omega}{2T}\bigg) \bigg].
\end{equation}
Integrating over the initial and final times of the stroke then yields the total work performed: 
\begin{equation}
\mathcal{W}_\text{iso} =  T \ln \bigg[\frac{\sinh(\Omega_f/2T)}{\sinh(\Omega_i/2T)}\bigg].
\end{equation}
In the classical limit $T \gg \Omega_{i,f}$ we get 
\begin{equation}
\mathcal{W}_\text{iso} \simeq T \ln \bigg( \frac{\Omega_f}{\Omega_i}\bigg).
\end{equation}

The total work performed in each stroke will therefore be 
\begin{IEEEeqnarray}{rCl}
\mathcal{W}_{ab} &=& T_H \ln \bigg[\frac{\sinh(\Omega_b/2T_H)}{\sinh(\Omega_a/2T_H)}\bigg],	\\[0.2cm]
\mathcal{W}_{bc} &=& \langle H \rangle_c - \langle H \rangle_b \\[0.2cm]
&=& \frac{\Omega_c}{2} \coth\bigg(\frac{\Omega_c}{2T_C}\bigg) - 
\frac{\Omega_b}{2} \coth\bigg(\frac{\Omega_b}{2T_H}\bigg),
\\[0.2cm]
\mathcal{W}_{cd} &=& T_C \ln \bigg[\frac{\sinh(\Omega_d/2T_C)}{\sinh(\Omega_c/2T_C)}\bigg],	\\[0.2cm]
\mathcal{W}_{da} &=& \langle H \rangle_a - \langle H \rangle_d \\[0.2cm]
&=& \frac{\Omega_a}{2} \coth\bigg(\frac{\Omega_a}{2T_H}\bigg) - 
\frac{\Omega_d}{2} \coth\bigg(\frac{\Omega_d}{2T_C}\bigg).
\end{IEEEeqnarray}
Moreover, the heat exchanged in $ab$ and $cd$ will be 
\begin{IEEEeqnarray}{rCl}
Q_{H} &=& \langle H \rangle_b - \langle H \rangle_a - \mathcal{W}_{ab},	\\[0.2cm]
Q_{C} &=& \langle H \rangle_d - \langle H \rangle_c - \mathcal{W}_{cd}.	
\end{IEEEeqnarray}

From this one may now compute the efficiency of the quasi-static cycle
\begin{equation}
\eta = - \frac{\mathcal{W}_{ab}+\mathcal{W}_{bc}+\mathcal{W}_{cd}+\mathcal{W}_{da}}{Q_H} = 1 + \frac{Q_C}{Q_H}.
\end{equation}
However, due to Eq.~(\ref{Carnot_constraint}) it follows that 
\begin{equation}
\frac{Q_C}{Q_H} = - \frac{T_C}{T_H},
\end{equation}
therefore leading us to the Carnot efficiency 
\begin{equation}
\eta_C = 1 - \frac{T_C}{T_H}.
\end{equation}
We emphasize that this result is only obtained if the constraint~(\ref{Carnot_constraint}) is applied.

\bibliography{/Users/gtlandi/Documents/library}

\end{document}